\documentclass[aps,prc,twocolumn,showpacs,superscriptaddress]{revtex4}
\usepackage[english]{babel}
\usepackage{amsmath}
\usepackage{amsfonts}
\usepackage{amssymb}
\usepackage{graphicx}

\begin{document}

\title{Excitation of $^{229}$Th nuclei in laser plasma:
the energy and half-life of the low-lying isomeric state}

\author{P.~V.~Borisyuk}
\email{pvborisyuk@mephi.ru (experiment)}

\affiliation{National Research Nuclear University MEPhI, 115409,
Kashirskoe shosse 31, Moscow, Russia}

\author{E.~V.~Chubunova}

\affiliation{National Research Nuclear University MEPhI, 115409,
Kashirskoe shosse 31, Moscow, Russia}

\author{N.~N.~Kolachevsky}

\affiliation{P.~N.~Lebedev Physical Institute of the Russian
Academy of Sciences, 119991, Leninskii prospekt 53, Moscow,
Russia}

\affiliation{National Research Nuclear University MEPhI, 115409,
Kashirskoe shosse 31, Moscow, Russia}

\author{Yu.~Yu.~Lebedinskii}

\affiliation{National Research Nuclear University MEPhI, 115409,
Kashirskoe shosse 31, Moscow, Russia}

\author{O.~S.~Vasiliev}

\affiliation{National Research Nuclear University MEPhI, 115409,
Kashirskoe shosse 31, Moscow, Russia}

\author{E.~V.~Tkalya}
\email{tkalya@srd.sinp.msu.ru (theory)}

\affiliation{Skobeltsyn Institute of Nuclear Physics Lomonosov
Moscow State University, Leninskie gory, Moscow 119991, Russia}

\affiliation{National Research Nuclear University MEPhI, 115409,
Kashirskoe shosse 31, Moscow, Russia}

\affiliation{Nuclear Safety Institute of RAS, Bol'shaya Tulskaya
52, Moscow 115191, Russia}

\date{\today}

\begin{abstract}
The results of experimental studies of the low-energy isomeric
state in the $^{229}$Th nucleus are presented. The work is
consisted of several stages. During the first stage $^{229}$Th
nuclei were excited with the inverse internal conversion to the
low-lying isomeric level in plasma that was formed by laser pulse
at the $^{229}$Th-containing target surface. Then thorium ions
having excited nuclei were extracted from the plasma by an
external electrical field and implanted into thin SiO$_2$ film
grown on a silicon substrate (that is a dielectric material with
about 9 eV band-gap). Gamma decay of isomeric nuclei was
registered during the second stage by the general methods of the
electron spectroscopy after the photon-electron emission from the
silicon substrate. Substitution of the photon registration with
the electron one allowed us to increase the desired signal by
several orders of magnitude and detect the $^{229}$Th nuclei
decay. During the third stage the electron spectra from standard
Xe VUV source was obtained that allowed determining the energy of
photons. In order to prove that the detected signal is caused by
isomeric $^{229}$Th nuclei decay a series of experiments was
carried. The analysis of electron spectra gives the following
results: the energy of the nuclear transition is
$E_{\text{is}}=7.1(^{+0.1}_{-0.2})$~eV, the half-life of the
isomeric level in bare nucleus in vacuum is
$T_{1/2}=1880\pm170$~s, the reduced probability of the isomeric
nuclear transition is $B_{\text{W.u.}}(M1;3/2^+\rightarrow
5/2^+)=(3.3\pm0.3)\times10^{-2}$.
\end{abstract}


\maketitle

\section{Introduction}

In 1976 Kroger and Reich discovered a low-lying state in
$^{229}$Th nucleus with the energy $E_{\rm is}<100$\,eV analyzing
$\gamma$-radiation after decay of $^{233}$U  isotope
$^{233}$U$\rightarrow^{229}$Th$+^{4}$He \cite{Kroger-76}. Further
experiments carried out by Reich and Helmer in 1990 indicated
the extremely low energy  of \mbox{$E_{\rm is}=1\pm4$\,eV} for the first
excited isomeric state \cite{Reich-90}. Four years later the same
authors  presented an improved value of $E_{\rm is}=3.5\pm1.0$\,eV
\cite{Helmer-94}. However, accurate measurements of
$\gamma$-transitions in the energy ranges of 29\,keV and 42\,keV,
made in 2007 by Beck~{\it et al.}  resulted in significantly
different value of $E_{\rm is}=7.6\pm0.5$\,eV \cite{Beck-07}
(later slightly corrected to  $7.8\pm 0.5$\,eV \cite{Beck-R}). The
most recent data coming from direct measurements of electron
internal conversion process predict that 6.3\,eV\,$\leq
E_{\text{is}}\,\leq \,18.3$\,eV~\cite{Wense-16}. In this work the
$\alpha$ decay process of $^{233}$U was used for populating the
low-lying isomeric state of $^{229}$Th  which occur with  2\%
probability \cite{Wense-16}. This method sets serious limitations
to available experimental configurations because of various decay
products, high background and relatively low particle number
\cite{Wense-16}. Direct excitation of the $^{229}$Th isomeric
nuclear state would significantly facilitate further research
activities, including variety of practical applications, but
remains one of the not yet resolved challenges.

Fundamental interest to this problem is based on a number of
unusual physical problems such as electron bridge
\cite{Strizhov-91}, isomeric state alpha decay \cite{Dykhne-96},
photon emission from the nucleus in a dielectric material with a
wide bandgap \cite{Tkalya-00-JETPL,Tkalya-00-PRC}, enhancement of
the relative effects of the variation of the fine structure
constant $\alpha$ and the strong interaction parameter
$m_q/\Lambda_{QCD}$ \cite{Flambaum-06} and others.

As for feasible  applications, the main focus is the development
of nuclear transition-based laser \cite{Tkalya-11,Tkalya-13}  and
highly accurate  optical clocks
\cite{Peik-03,Kazakov-12,Peik-15,Campbell-12,Borisyuk-17-QE,Wense-18}.
One of the approaches to nuclear transition-based optical clocks
is to laser cool a cloud of $^{229}$Th ions loaded in a Paul
trap~\cite{Herrera-Sancho-13,Herrera-Sancho-12,Okhapkin-15}.
Similar approach was developed for trapping $^{229}$Th$^{3+}$
Coulomb ion crystals, the first successful trapping of $10^{4}$
ions was reported in \cite{Radnaev-12}. In these experiments
electronic transitions with energies near 7.8 eV  in trapped
thorium ions were carefully measured by laser spectroscopy. The
uncertainty budget of proposed $^{229}$Th optical clocks was
analyzed in \cite{Campbell-12}. Taking into account the Zeeman and
Stark shifts, the Doppler effect, black body radiation shift,  and
ion micromotion, the relative frequency uncertainty was  estimated
as $<10^{-19}$. This is significantly lower than for the best
known state-of-the art optical clocks based on electronic
transitions \cite{Takano-16}.

One of the approaches to directly excite  the low-lying isomeric
state  with photons was developed by  Jeet and co-authors
\cite{Jeet-15}, and by Stellmer and co-authors \cite{Stellmer-18}.
$^{229}$Th ions were implanted in transparent crystals like
LiCaAlF$_6$, LiSrAlF$_6$, CaF$_2$ and other samples
\cite{Rellergert-10,Hehlen-13,Dessovic-14,Stellmer-15}) and then
illuminated by intensive  VUV (vacuum ultra violet)  radiation.
These  experiments did not succeed in excitation of the isomeric
state.  One of the feasible reasons could be extremely broad
energy range (of approximately one eV or 250\,THz) needed to scan
in search for the resonance.  For any optical method like
synchrotron radiation spectroscopy or laser spectroscopy it will
require incredibly long time to cover this spectral range, taking
into account a narrow expected resonance line width.

We study excitation of $^{229}$Th nuclei by inverse internal
conversion process in laser plasma which was suggested in
\cite{Strizhov-91}. The inverse internal conversion process (IIC
\cite{IIC-NEEC}) was discussed for the first time by Goldanskii
and Namiot in \cite{Goldanskii-76} and analyzed in details in the
later work \cite{Tkalya-04}. After ablation of $^{229}$Th target,
ions can be implanted into a wide-gap dielectric material
(SiO$_2$) to prevent electronic internal conversion process which
reduces the lifetime of the nuclear isomeric state
\cite{Tkalya-00-JETPL,Tkalya-00-PRC}. Combination of these two
methods (plasma formation and ion implantation) allow to directly
excite nuclear isomeric state and measure its energy and the
lifetime.

\section{Excitation $^{229}\textrm{Th}$ nuclei in laser plasma}
\label{sec:ICC}

\begin{figure}
\includegraphics[angle=0,width=0.95\hsize,keepaspectratio]{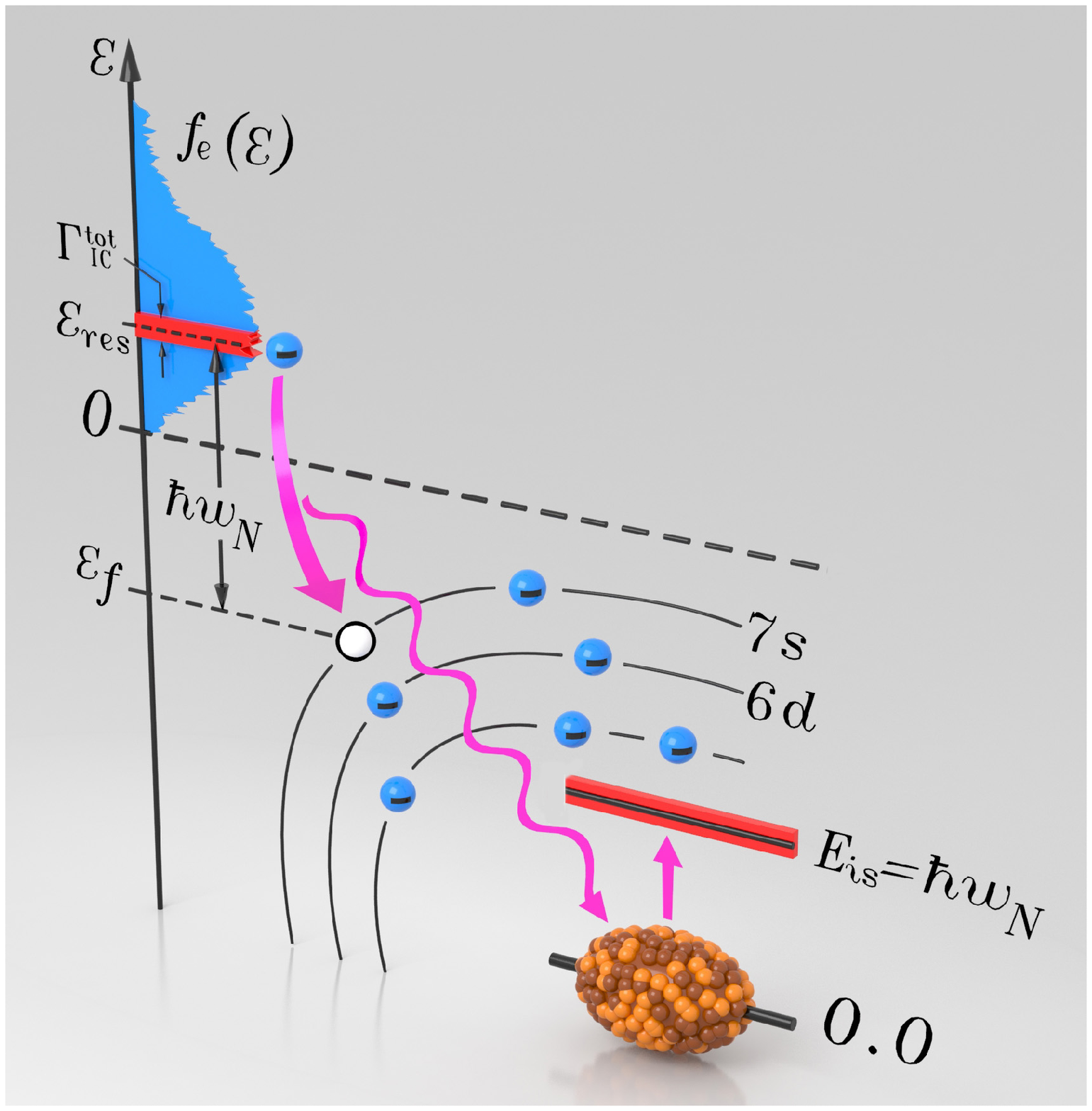}
\caption{Excitation of $^{229}$Th nucleus via inverse internal
conversion to the 7$s$ electron shell.}
\label{fig:ICC}
\end{figure}

Laser ablation from $^{229}$Th-containing target is one of the
methods to directly excite isomeric nuclear state. Theoretical and
experimental studies show (see in \cite{Tkalya-04} and references
therein) that the inverse internal conversion process is the most
efficient for nuclei excitation in the case of $^{229}$Th. Plasma
electrons of the continuous spectrum with energies ${\cal{E}}$
contribute to IIC process and occupy the ion states of discrete
spectrum with energies ${\cal{E}}_f$. The nucleus in this case is
excited by a virtual photon (see in Fig.~\ref{fig:ICC}).
%
%
The considered process is inverse  to the isomeric nuclear state
decay via internal electron conversion. IIC cross-section for
completely vacant ion shell is calculated using the conventional
methods of the perturbation theory of quantum electrodynamics
\cite{Tkalya-04}
\begin{equation}
\sigma_{\textrm{IIC}}({\cal{E}}\rightarrow{}{\cal{E}}_f)\simeq
\delta(E_{\text{is}}-{\cal{E}}+{\cal{E}}_f)
\frac{\lambda_e^2}{4}\Gamma_{\text{IC}}(E_{\textrm{is}},f)g ,
\label{eq:Cross-section}
\end{equation}
where $\delta$ is the Dirac delta-function, $E_{\textrm{is}}$ is
the energy of the nuclear isomeric state,
$\Gamma_{\text{IC}}(\omega_N,f)$ is a partial internal conversion
(IC) width of isomeric state for the decay via $f$ ion shell,
$\lambda_e$ is the de Brogile wavelength of plasma electron,
$g=(2J_F+1)/(2J_I+1)$, and $J_{I,F}$ are the nuclear spins in
initial ($I$) and final ($F$) states.

Excitation is caused by electrons in the plasma spectrum with
energies ${\cal{E}}_{\text{res}}$ approaching (within the width
$\Gamma_{\text{IC}}$) the difference of the nuclear transition
energy $\omega_N$ and the binding energy $|{\cal{E}}_f|$ of the
occupying atomic shell. I.e. the process has a pronounced
resonance character. The IIC process is most efficient at the
electron plasma temperature $kT$ comparable to $\hbar\omega_N$ (in
our case $kT\simeq \hbar\omega_N = E_{\text{is}}$). At such
temperatures, first, the atomic shells contributing to IIC are
ionized, and second, the electron density with the energies around
$\hbar\omega_N$ is high enough. To reach temperature corresponding
to the energy of $\approx10$\,eV, a moderate laser intensity of
$I\sim 10^{10}$ W\,cm$^{-2}$ is required \cite{Torrisi-06}.

The probability to excite isomeric state in plasma with  electron
density $n_e$ and electronic energy distribution function
$f_e({\cal{E}})$ is given by \cite{Tkalya-04}
\begin{equation}
\eta_{\textrm{IIC}} = \int_0^{\infty}
\sigma_{\textrm{IIC}}({\cal{E}}\rightarrow{}{\cal{E}}_f) n_e
f_e({\cal{E}}) \tau \upsilon_e \frac{d{\cal{E}}}{kT}\,,
\label{eq:Efficiency-Int}
\end{equation}
where $\upsilon_e=\sqrt{2{\cal{E}}/mc^2}$ is  electron velocity
($m$ is the electron mass), and $\tau$  is a duration of
excitation process or typical laser plasma life time.

After integration the delta function on energy from
$\sigma_{\textrm{IIC}}$ is eliminated. Then the probability can be
evaluated as
\begin{equation}
\eta_{\textrm{IIC}} \simeq \sigma_{\textrm{IIC}}^{\textrm{eff}}
f_e({\cal{E}}_{\textrm{res}}) n_e \tau \upsilon_e^{\textrm{res}},
\label{eq:Efficiency}
\end{equation}
where
\begin{equation}
\sigma_{\textrm{IIC}}^{\textrm{eff}}=
\left(\frac{\lambda_e^{\text{res}}}{2}\right)^2
\frac{\Gamma_{\text{IC}}^{\textrm{tot}}}{kT} \simeq 2\times
10^{-25}\,\, \text{cm}^2
\end{equation}
is the effective cross-section,
$\Gamma_{\text{IC}}^{\textrm{tot}}\simeq 3\times 10^{-10}$~eV is
the total IC width of the nuclear level [this preliminary estimate
for the conversion width follows from the middle value of the
reduced probability of the nuclear isomeric transition,
$B_{W.u.}(M1;3/2^+\rightarrow 5/2^+)\simeq 3.1\times 10^{-2}$ (see
Table~I in \cite{Tkalya-15-PRC}), and the internal conversion
coefficient $\alpha_{M1}\simeq 1.5\times10^{9}$ (the result of
calculation by the computer code \cite{Band-79} for the transition
energy of 8~eV)], ${\cal{E}}_{\text{res}}$ is the electron
resonance energy can be evaluated as
${\cal{E}}_{\text{res}}=E_{\text{is}}-|{\cal{E}}_f| \simeq 1$~eV,
since for $7$s-electrons the binding energy in thorium atom equals
${\cal{E}}_f=6.3$\,eV \cite{NIST-15}, parameter
$\lambda_e^{\text{res}}=2\pi\hbar/\sqrt{2m{\cal{E}}_{\text{res}}}$
corresponds to the resonance de Broglie wavelength, $n_e$ is the
electron density in plasma, and $\tau$ is the plasma's life time
which equals to the duration of the laser pulse
$\tau_{\text{L}}\approx 15$~ns (see below).

The number of the $^{229}$Th nuclei evaporated by a laser pulse
from the target is
\begin{equation}
N_{229}^{\text{evap}} \simeq
\frac{\beta\,\rho_{\text{ThO}_2}\,V}{[(1-\beta)(232+32)+
\beta(229+32)]u} \simeq 2\times 10^{13}\,,
\end{equation}
where $\rho_{\text{ThO}_2}=10$~g\,cm$^{-3}$ is the density of the
thorium target, $u=1.66\times 10^{-24}$~g is the atomic mass unit.
Here $V=\pi{}w_L^2 h$ is the laser plasma volume with the laser
spot waist of $w_L=5\times10^{-3}$\,cm, and typical penetration
depth of laser radiation of $h\simeq 10^{-4}$\,cm. The latter is
defined by the wavelength of excitation laser
($1.06\times10^{-4}$\,cm). Parameter  $\beta=0.068$  stands for
the fraction of $^{229}$Th isotope in the target. In our
experiments we used target containing 6.8\% of the $^{229}$Th
nuclei with the rest of $^{232}$Th nuclei (93.2\%).

After evaporation of the  target, the plasma expands. The front of
the shock wave, which contains approximately of 10\% of the
evaporated $^{229}$Th nuclei \mbox{$N_{229}\simeq
0.1N_{229}^{\text{evap}}$}, has a number density of thorium ions
of \mbox{$\simeq 10^{20}$~cm$^{-3}$} and a number density of
electrons \mbox{$n_e\simeq 10^{20}$~cm$^{-3}$} (the plasma is
neutral as a whole). The front of the wave moves with the speed of
sound absorbing laser radiation for a time $\tau_{\text{L}}$ and
warming up to the electronic temperature of $kT\simeq 10$~eV.

Thus, for Maxwellian electron energy distribution $f_e({\cal{E}})$
the excitation probability per one nucleus of $^{229}$Th can be
evaluated from Eq.~(\ref{eq:Efficiency}). The result is
\begin{equation}
\eta_{\textrm{IIC}} \simeq 10^{-5}.
\label{eq:Efficiency-Number}
\end{equation}
Finally, we can evaluate the number of excited nuclei per laser
pulse as
$$
N_{\text{is}}\simeq \eta_{\textrm{IIC}} N_{229} \simeq10^8,
$$
which sounds promising for further analysis. Besides excitation of $^{229}$Th
nuclei in laser ablation, this method allows to implant ionized
thorium into a wide bandgap dielectric matrix and analyse them {\it in situ}.

\section{Ion implantation into $\textrm{SiO}_2$ matrix}
\label{sec:ionimp}

\begin{figure}
\includegraphics[angle=0,width=0.90\hsize,keepaspectratio]{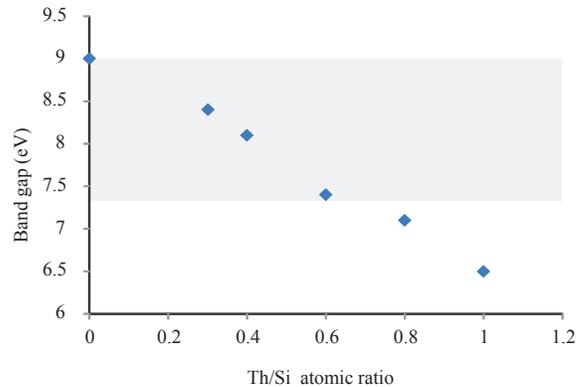}
\caption{Dependency of the $^{232}$Th:$\textrm{Si}\textrm{O}_2$
compound band gap on relative number of thorium atoms according to
surface chemical composition analysis using XPS and REELS methods
\cite{Borisyuk-18-LPL}.} \label{fig:Th:SiO_2BandGap}
\end{figure}

Previous studies of laser implantation of $^{232}$Th ions in
SiO$_2$ matrix \cite{Borisyuk-18-LPL} showed that a wide band
dielectric compound $^{232}$Th:$\textrm{Si}\textrm{O}_2$ is formed
in the subsurface area. Dependency of the band gap on Th/Si atomic
fraction is shown in Fig.~\ref{fig:Th:SiO_2BandGap}. The large
band gap of at least 8~eV~\cite{Beck-R} is necessary to prevent
prompt IC process of excited $^{229}$Th nuclei. In this case
(marked with gray color) one can expect that implanted excited
isomeric $^{229}$Th nuclei will slowly decay via $\gamma$ channel.
To reach this regime,  the atomic ratio Th/Si should be less than
0.4.
%
%

\begin{figure}
\includegraphics[angle=0,width=0.90\hsize,keepaspectratio]{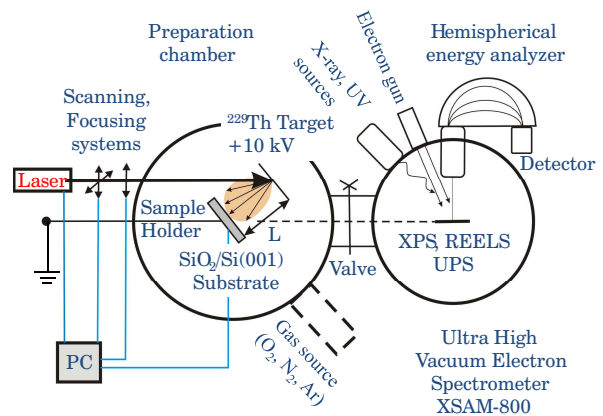}
\caption{Experimental setup for target preparation and \textit{in
situ}  analysis based on electron spectrometer XSAM- 800 Kratos. 1
--- sample loading and target preparation chamber, 2 --- YAG:Nd
laser for target ablation, 3 --- precision manipulator 4 ---
valve, 5 --- analysis chamber, 6 --- hemispherical electron
analyzer chamber equipped by X-ray and UV sources, and e-gun.}
\label{fig:ExpSetup}
\end{figure}

To provide this condition and to prepare the sample, we use method
depicted in Fig.~\ref{fig:ExpSetup}. The experimental setup bases
on the XSAM-800 (Kratos, UK) electron spectrometer with the sample
loading chamber upgraded for ablation and implantation of Th ions.
The method allows to  study different samples  \textit{in situ} in
the ultra high vacuum (UHV) surface analysis system. Implantation
of Th ions in SiO$_{2}$/Si target was performed directly in the
spectrometer's preparation chamber.
%
%
For laser ablation we use a target of thorium oxide (thin disk of
3\,mm in diameter)  deposited on 10\,mm$\times$10\,mm silicon
substrate. The overall activity of the $^{229}$Th isotope in the
target is around  $100$ kBq which  is equivalent to $10 ^{16}$
nuclei. The main source of radioactivity are $^{229}$Th nuclei
which provide 99.988\% of all $\alpha$ emitters in the target.
Another source are  $^{228}$Th nuclei, which contribute by
approximately 0.011\%.  The isotope ratio of $^{229}$Th in the
target equals 6.8\%, the rest are $^{232}$Th nuclei.

Focused radiation of a solid state Nd:YAG laser at
$\lambda=1064$\,nm wavelength is used for target ablation. The
laser operates in the  Q-switched regime with the pulse energy of
100 mJ and pulse duration of 15 ns  which provides  power density
$I$ on a target in the range $10^9-10^{10}$\,W/cm$^2$. Laser power
density is the most critical parameter defining excitation
probability $\eta_\textrm{IIC}$. It determines electron
temperature, degree of ionization, particle number density which
enter the expression for
$\eta_\textrm{IIC}$~(\ref{eq:Efficiency-Int}). The laser focus was
scanned on a target using deflection prisms controlled by a
personal computer (PC). The amount of sputtered material could be
varied by changing power density and the number of laser pulses.
Ion component of laser plasma was accelerated in an electrical
field between the target and SiO$_2$/Si substrate placed parallel
to the target. Before the laser shot, the target has a potential
of +10\,kV  and is connected to the grounded substrate by 25\,nF
capacitor. During the laser shot, the capacitor discharges through
plasma, while ions are accelerated and hit the substrate with
energies of 10\,keV which is enough to implant in SiO$_2$
dielectric. The charge accumulated in the capacitor before the
laser shot was adjusted to the expected net charge of the ion
component in plasma  (typically $\sim10^{12}-10^{13}$ ions per
shot with the degree of ionization $\simeq50$\%).

The target and the sample are separated by the distance of
$L=5$\,cm. We evaluate the time-of-flight of thorium ion between
the target and the sample as $<10^{-7}$\,s. It means, that
independently of ion charge multiplicity, excited nuclei reach the
target before decay via IC or other channels (see below). The
large band gap of SiO$_2$ will block IC process immediately after
implantation.

For sample preparation we typically use 5 laser pulses with the
repetition rate of 0.2\,Hz. Then the  Th:SiO$_2$/Si sample was
moved under UHV conditions from the preparation chamber into the
analysis chamber using a mechanical tool. The transportation time
from one chamber to another chamber is  less than 30\,s. In most
experiments we use 6.5\,nm SiO$_2$ film on Si(001) crystal,
$p$-type 0.05\,Ohm\,cm. Such thickness provides optimal
sensitivity for photoelectron spectroscopy method which is
described in the next Section. Concentration of thorium in the
SiO$_2$ film was measured by X-ray photoelectron spectroscopy
(XPS) with Al\,$K_{\alpha}$ emission at $1486.6$\,eV. Fractional
number density of implanted Th nuclei is less than 1\% which,
according to fig.\,\ref{fig:Th:SiO_2BandGap}, corresponds to the
Th:SiO$_2$ band gap of 9.0\,eV. The band gap was measured by
reflected electron energy loss spectroscopy (REELS). 

\section{Electron spectroscopy}
\label{sec:Electron_spectroscopy}

\begin{figure}
\includegraphics[angle=0,width=0.95\hsize,keepaspectratio]{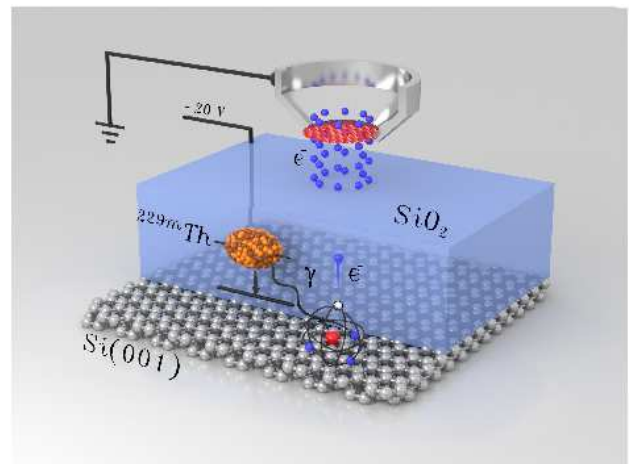}
\caption{Decay of an excited $^{229}$Th nucleus followed by
emission of photoelectron from Si substrate. After penetration
through 6.5\,nm SiO$_{2}$ film photoelectons  are collected by
XSAM- 800 electron spectrometer.}
\label{fig:229mThDecay-Si-Electron}
\end{figure}

We study  $^{229}$Th:SiO$_2$/Si sample using  the sensitive
XSAM-800 electron spectrometer (Fig.~\ref{fig:ExpSetup}). We
expect, that VUV photons emitted in $\gamma$ decay of excited
$^{229}$Th nuclei interact with Si substrate which plays a role of
a photocathode. If SiO$_2$ layer is thin enough, photoelectrons
can leave the sample and reach the spectrometer. The process is
schematically shown in Fig.~\ref{fig:229mThDecay-Si-Electron}.

%
%
Thickness of silicon oxide used in experiments is determined, from
one side, by implantation depth of thorium ions (typically $10$
nm) and, from the other side, by extinction of electron signal in
the SiO$_2$ layer. In most of experiments we used 6.5\,nm thick
SiO$_2$ film which suppresses electron signal by approximately 6
times for a normal incidence electron. To prevent cut-off close to
zero photoelectron kinetic energies the accelerating potential of
$-20.0$\,V relative to the entrance aperture of the spectrometer
is applied. Thus, the sample itself plays a role of an efficient
photo cathode converting $\gamma$-quanta to low energy electrons.
Compared to direct registration of $\gamma$-quanta (VUV photons)
this approach provides much higher sensitivity.

\begin{figure}
\includegraphics[angle=0,width=0.95\hsize,keepaspectratio]{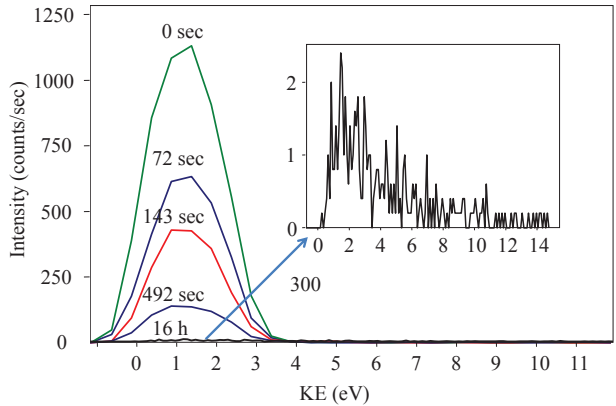}
\caption{A sequence of electron  energy spectra measured after
different time intervals from the measurement start (shown on the
plot).  The inset shows the spectrum measured in 16 hours after
implantation (magnified). Horizontal axis is corrected for the
$20.0$\,eV offset corresponding to potential difference between
the sample and the spectrometer's aperture
(Fig.~\ref{fig:229mThDecay-Si-Electron}). KE is photoelectron
kinetic energy.} \label{fig:ElectronEnergySpectraThroughSiO2}
\end{figure}

Fig.~\ref{fig:ElectronEnergySpectraThroughSiO2} shows a sequence
of electron energy spectra measured after different time intervals
counted from the measurement start. For clarity, the spectra are
shifted by~20.0~eV corresponding to potential difference between
the sample and the spectrometer's aperture. Strong signal (up to
3000\,counts\,s$^{-1}$) is observed in the energy interval from
0\,eV to 3\,eV. The signal decreases in time and vanishes in
several hours. Note, that after long time interval (e.g.
16\,hours) the spectrum transforms in a  background of
approximately 15\,eV width and integrated count rate of $\simeq
7$\,s$^{-1}$ (see the inset in
Fig.~\ref{fig:ElectronEnergySpectraThroughSiO2}). Knowing the
intensity of the background signal and half-life of the
radionuclide, one can evaluate the total number of implanted
$^{229}$Th nuclei as $3\times 10^{12}$. Within measurement
uncertainty this coincides with estimate of thorium atomic
concentration from XPS measurements.

%
%

We performed a set of experiments with different ablated targets
to study origin of the narrow-line signal shown in
Fig.~\ref{fig:ElectronEnergySpectraThroughSiO2}. Although there
are no known long-living electronic states in silicon or silicon
oxide which can be excited during ion implantation, we did
experiments with pure silicon, pure carbon and some other targets
for laser ablation (for more details see
Section\,\ref{sec:appI} (Appendix)). After ablation, the procedure of sample
preparation and analysis described in the Section \ref{sec:ionimp}
was reproduced. In these experiments we did not detect any signal
in the expected  energy range (0--3 eV). This observation confirms
the absence of long-living electronic states in the sample. Even
more stringent test was done with targets of isotopically pure
$^{232}$Th and $^{232}$Th oxide film. We also did not see any
signal similar to the one shown in
Fig.~\ref{fig:ElectronEnergySpectraThroughSiO2}, independently on
the laser power density $I$. It proves that the origin of the
signal comes purely from $^{229}$Th atoms and we ascribe this
peculiar signal to the process depicted in
Fig.~\ref{fig:229mThDecay-Si-Electron}. We should note, that the
signal amplitude is very sensitive to the laser power density $I$,
which also confirms that the signal results from  ICC process
described in the Section\,\ref{sec:ICC}. Verification of results
is described in more details in Section\,\ref{sec:appI} (Appendix).

\section{The nuclear transition energy from calibrated electron spectra}

\begin{figure}
    \includegraphics[angle=0,width=0.95\hsize,keepaspectratio]{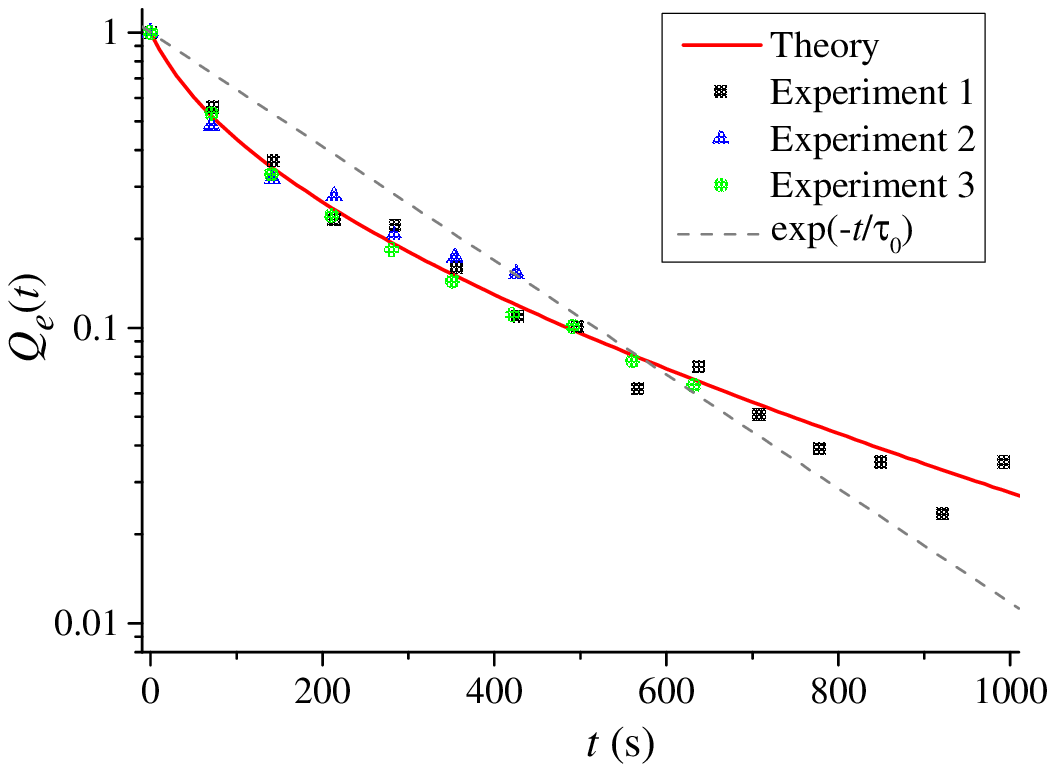}
    \caption{Time dependency of normalized experimental signal (dots)
        from three independent series of measurements.
        Dashed line --- exponential decay with the constant
        $\tau_0=225$\,s, red line --- theoretical model from equation
        (\ref{eq:Q_e}) for  energy $E_{\text{is}}=7.1$~eV and
        $T_{1/2}=1880$ s.}
    \label{fig:Q_e(t)}
\end{figure}

Good reproducibility and strong  signal from $^{229}$Th:SiO$_2$/Si
sample allows to accurately measure dynamics of the process.
Figure~\ref{fig:Q_e(t)} shows time dependency of normalized signal
obtained in three independent experiment series (in a logarithmic
scale). The broad background coming from $\alpha$-decay of
$^{229}$Th nuclei (the inset in
Fig.~\ref{fig:ElectronEnergySpectraThroughSiO2}) is subtracted
from the spectra before evaluation the  data for
Fig.~\ref{fig:Q_e(t)}. The decay deviates from exponential which
can be interpreted as the Purcell effect (change of decay
probability due to environment~\cite{Purcell-46}) which is
described in details in the Section \ref{sec:T-B}.
%
%
%

\begin{figure}
\includegraphics[angle=0,width=0.99\hsize,keepaspectratio]{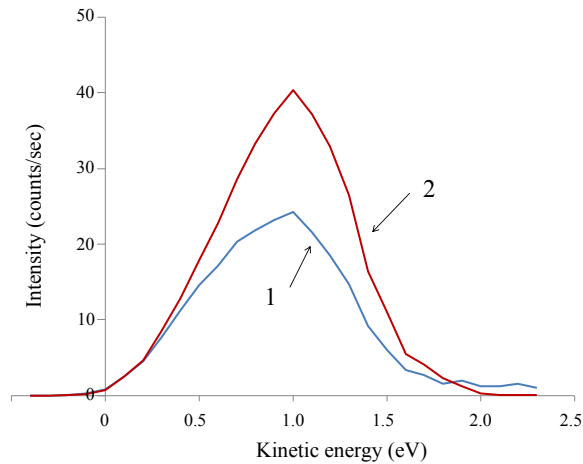}
\caption{High resolution photo electron spectrum from
$^{229}$Th:Si0$_2$ target measured during 500\,s (line\,1). The
spectrum shape is corrected according to the decay curve of
Fig.~\ref{fig:Q_e(t)}, then the  the background coming from
$\alpha$-decay of $^{229}$Th is subtracted. Line\,2 shows the
corrected spectrum. Calibration of the energy scale was carried
out using Si2$p$ line form silicon substrate obtained during XPS
measurements (AlK$\alpha$ X-ray source was used).}
\label{fig:ElectronSpectraVUVLamps}
\end{figure}

To recover the energy of VUV photons emitted during decay of
isomeric nuclei from photo electron spectra, we performed high
resolution  measurements. To achieve the uncertainty of 0.1~eV the
spectra with high count rate of several thousand of electrons per
energy bin are demanded, which requires long acquisition time of a
few minutes. High-resolution spectrum measured during 500\,s is
presented in Fig.~\ref{fig:ElectronSpectraVUVLamps}, line\,1.
Reduction of the signal during the measurement results in
distortion of the spectrum. We corrected the measured spectra
according to the decay curve of  Fig.\,\ref{fig:Q_e(t)} and then
subtract the background coming from $\alpha$-decay of $^{229}$Th
nuclei  (the inset in
Fig.~\ref{fig:ElectronEnergySpectraThroughSiO2}). The corrected
spectrum is shown in Fig.~\ref{fig:ElectronSpectraVUVLamps},
line\,2.

%
%

The energy $\hbar\omega_N$  of the emitted $\gamma$-quantum during
isomeric transition in the $^{229}$Th nucleus can be derived from
usual equation of the photoeffect:
\begin{equation*}\label{eq:KE}
KE=\hbar\omega_N-BE-eU.
\end{equation*}
Here $KE$ is the kinetic energy of electrons emitted from the
silicon substrate (measured value), $BE$ is the electron binding
energy relatively to the vacuum energy level ($E_{\text{vac}}$),
$U$ is the potential difference between the sample and the
detector. The latter is equal to the difference between the work
function of the sample and that of the detector
($eU=WF_{\textrm{d}}-WF_{\textrm{Si}}$).
In the case when $WF_{\textrm{Si}}<WF_{\textrm{d}}$, a part of the
spectrum with electron energies close to zero is lost. To prevent
this effect in our measurements, a negative potential
$U_0=-20.0$\,V is applied to the sample, which gives
$eU=eU_0+WF_{\textrm{d}}-WF_{\textrm{Si}}$.
This equation consists the term $WF_{\textrm{d}}$ which value
differs for different devices and experimental conditions. In
order to determine it one can use calibration of the energy scale
of the analyzer using core levels of a reference material,
obtained during XPS measurements. In our case, AlK$\alpha$
emission line with energy 1486.6~eV was used and a silicon was
chosen to be a reference material as far as it's core lines are
defined well. At first, such calibration allows one to shift the
kinetic energy scale by choosing the $WF_{\textrm{d}}$  value in
order to make the position of spectral lines corresponded the $KE$
values counted from the vacuum level of silicon
substrate~\cite{Fujimura-2016}. Secondly, it allows one to obtain
the precision of kinetic energy determination about $\pm$0.05~eV
over the whole energy scale due to the narrowness of XPS core
lines \cite{Briggs-2003}.

%
%
\begin{figure}
\includegraphics[angle=0,width=0.99\hsize,keepaspectratio]{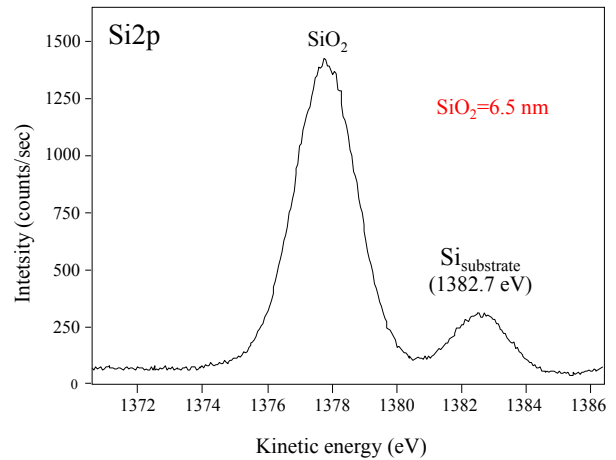}
\caption{Calibration XPS spectra of Si2$p$ lines for the sample
Th:SiO$_2$/Si formed after Th implantation.}
\label{XPS_Si2p_Th:SiO2/Si}
\end{figure}

XPS spectra of the sample under study after Th implantation is
presented in Fig.~\ref{XPS_Si2p_Th:SiO2/Si}. The energy difference
between the Si2$p$ core level and the Si valence band maximum
(VBM) reported in the literature is 98.80~eV \cite{Perego-12}.
Thus, the Si2p binding energy relatively to the vacuum level for
pure silicon substrate must be
\mbox{$BE_{\text{Si2p}}$=98.8+5.1=103.9~eV}. Converting this value
to the values for kinetic energies of photoelectrons, knocking out
from the level by X-ray AlK$\alpha$ radiation one can obtain  the
value \mbox{$KE_{\text{Si2p}}$=1486.6-103.9=1382.7~eV}. The
position of the Si2p spectral line corresponding to the signal
from silicon substrate in Fig.~\ref{XPS_Si2p_Th:SiO2/Si} is
connected to that value. The use of such energy scale calibration
independently on the applied voltage between the sample and the
spectrometer allows one to count the $KE$ value from the vacuum
level of silicon substrate. This calibration was applied to the
energy scale in Fig.~\ref{fig:ElectronSpectraVUVLamps} thus the
right part of the spectrum with maximal value of kinetic energy
($KE_{\text{max}}$) would correspond to the energy of
photoelectrons  from the edge of the Si valence band:
\begin{equation}\label{key}
KE_{\textrm{max}}=\hbar\omega_N -WF_{\textrm{Si}}.
\end{equation}
The left edge of the spectrum ($KE_{\text{min}}$) corresponds to
photoelectrons which leave the surface at energies close to zero.
They can be  not only primary low-energy electrons which are
directly exited from a filled state with an energy below
$E_{\text{vac}}$, but also higher  energy electrons scattered on
their way toward the surface. These electrons define the cutoff of
a photoelectron spectrum and define the zero of the energy scale
for the case of clean silicon.
However, in our case there is a  dielectric SiO$_2$ layer on Si
substrate, which means that  the measured $KE_{\text{min}}$ should
be equal to the potential difference between SiO$_2$ surface end
detector. Due to the presence of a dipole at SiO$_2$/Si interface,
the potential of the SiO$_2$ surface is different from the one of
Si. It means, that the photoelectron spectrum coming from the Si
substrate can change significantly in the cutoff energy  region
after  passing through the energy barrier at the interface.

In order to determine $\hbar\omega_N$  it is necessary to
determine the absolute $KE_{\text{max}}$  value after the above
mentioned spectrum energy scale calibration procedure. However,
the shape of the electron spectrum caused by UV sample irradiation
has a complex structure. In the easiest model if not taken into
account the energy dependence of such important physical
parameters like density of electronic states, photoionization
cross-section of valence electrons, the values of matrix elements
of electron transitions between sublevels of the valence band and
conductance band, the spectrum is mainly determined by the
processes of elastic and inelastic scattering.

Really, according to Ref.~\onlinecite{Berglund-1964} the main part
of photoelectrons, before leaving the sample encounter several
acts of inelastic collisions, as a result of which the initial
photoelectron lose part of it's initial energy. It leads to the
fact that a part of a spectrum formed by electrons with maximal
energy is redistributed to the spectrum energy range corresponding
to low electron kinetic energies. Thus the main point is shifted
to the minimal kinetic energy. At the same time, the right edge of
the spectrum has a monotonically damped character and there can be
no sharp edge of the spectrum observed usually for metals while
irradiating by UV of 20 or 40~eV or X-ray. The modeling of the
spectral line shape in our case is a complex task, that requires
determination of many parameters, including the ones related to
device. In order to avoid labour-intensive and probably non
correct model calculations within the framework of this work we
use a simple empirical method that is based in the comparison of
the right edge of the photoelectron spectra obtained for the
studied samples and VUV sources with known spectral
characteristics.

\begin{table}[b]
    \caption{Spectral lines of xenon discharge lamp and
        corresponding $KE_{\textrm{max}}$ energies for the photoelectron
        spectra presented in
        Fig.\,\ref{fig:Xe-KrPhotoelectronSpectraFromSi}. The last raw
        shows deduced $KE_{\textrm{max}}$ energy for $^{229}$Th:SiO$_2$/Si
        sample with decaying isomeric $^{229}$Th nuclei.}
    \label{tab:Lamps}
    \begin{tabular}{c|c|c|c}
        \hline
        UV Source & $\lambda$ (nm) & $\hbar\omega$ (eV) & $KE_{\text{max}}=\hbar\omega-WF_{\text{Si}}$ (eV)\\
        \hline
        \hline
        Xe & 147  & 8.4  &  $3.3(^{+0.1}_{-0.2})$    \\
        Th:SiO$_2$/Si &        &  $7.1(^{+0.1}_{-0.2})$     & $2.0(^{+0.1}_{-0.2})$  \\
        \hline
    \end{tabular}
\end{table}

\begin{figure}
\includegraphics[angle=0,width=0.99\hsize,keepaspectratio]{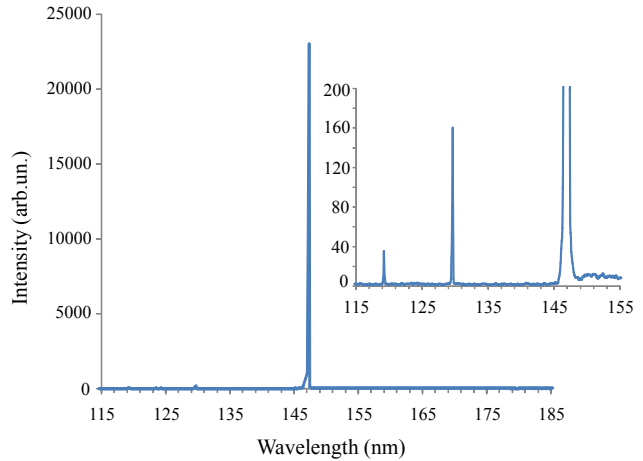}
\caption{VUV spectra of xenon (Xe) discharge lamp manufactured by
CHROMDET Ltd. (Russian Federation).}
\label{fig:VUVspectraXe}
\end{figure}

\begin{figure}
\includegraphics[angle=0,width=0.99\hsize,keepaspectratio]{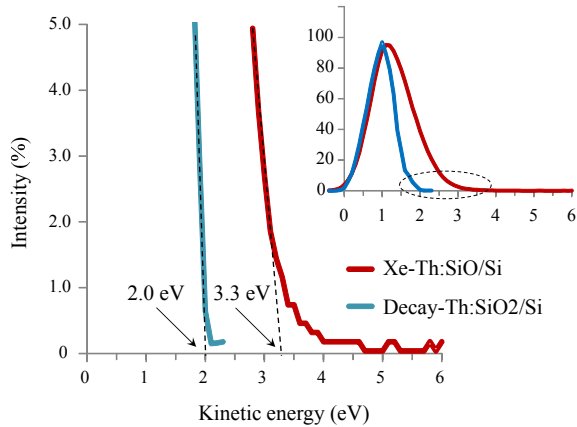}
\caption{Normalized photoelectron spectra from
$^{229}$Th:SiO$_2$/Si sample after illumination by VUV radiation
of xenon (Xe) discharge lamps. The same plot shows the
photoelectron spectrum from $^{229}$Th:SiO$_2$/Si sample during
isomeric decay. Maximal electron energy $KE_{\textrm{max}}$ for
each  spectrum is derived from extrapolation of the linear
function tangent to the right wing of the spectrum to the x-axis.}
    \label{fig:Xe-KrPhotoelectronSpectraFromSi}
\end{figure}

We used UV source (Xe discharge lamps) for modeling
photoionization from Th:SiO$_2$/Si. The photon energy of this lamp
is close to the expected spectral line coming from the $^{229}$Th
isomeric state decay (see Table~\ref{tab:Lamps}). The emission
spectra of this lamp (see Fig.~\ref{fig:VUVspectraXe}) together
with corresponding photoelectron spectra from decay Th:SiO$_2$/Si
are analyzed in details. Using results of our analysis, we agreed
of the following procedure for preparation of photoelectron
spectra:
\begin{itemize}
    \item after implantation, we recorded  photoelectron spectra from
$^{229}$Th:SiO$_2$/Si sample in high resolution mode. The sample
substrate is at \mbox{$U_0=-20.0$}\,V potential corresponding to
the spectrometer aperture. The spectra intensity are corrected
according to procedure described in
Fig.~\ref{fig:ElectronSpectraVUVLamps};
    \item using the same high
resolution mode and the same $U_0$ applied to the sample we
recorded  photoelectron spectra from  $^{229}$Th:SiO$_2$/Si sample
under illumination by Xe VUV lamp and under X-Ray souse
AlK$\alpha$ (1486.6 eV);
    \item for all spectra we use the same calibration of the energy scale,
namely, shift of the scale to the position of the Si2$p$ line,
which is obtained using the X-Ray source AlK$\alpha$ at
$KE_{\text{Si}2p}$=1382.7 eV;
    \item the analysis of the spectrum
of sample irradiated with Xe VUV lamp showed that the right values
of $KE_{\text{max}}$ is obtained when using the extrapolation with
linear function tangent to the high-energy shoulder of the
spectrum in the range of spectrum intensity from 5\textsl{\%} to
1\textsl{\%} and finding the intersection with x-axis as shown in
Fig.~\ref{fig:Xe-KrPhotoelectronSpectraFromSi}. The accuracy of
determining $KE_{\text{max}}$ by this method has asymmetry
character and better than -0.2/+0.1~eV;
    \item the same
method definition of $KE_{\text{max}}$  we use for spectrum,
obtained for  $^{229}$Th:SiO$_2$/Si sample during decay
(Fig.~\ref{fig:Xe-KrPhotoelectronSpectraFromSi}).
\end{itemize}

%
%

Fig.~\ref{fig:Xe-KrPhotoelectronSpectraFromSi} shows high
resolution photoelectron spectra from  $^{229}$Th:SiO2/Si sample
irradiated by VUV lamp and during isomeric decay after intensity
normalization and energy calibration. Using the same method of
tangent line we derive that the  maximum kinetic energy of
photoelectrons $KE_{\textrm{max}}$ excited in silicon by photons
during $^{229}$Th isomeric state decay  in Th:SiO$_2$/Si structure
is $2.0(^{+0.1}_{-0.2})$ eV. The uncertainty comes from combined
fit and extrapolation uncertainties of $KE_{\textrm{max}}$ and the
energy calibration uncertainty  of the electron spectrometer
according to its specification. Results are summarized in
Table~\ref{tab:Lamps}.

%
%

Thus, using the value $KE_{\textrm{max}}=2.0(^{+0.1}_{-0.2})$ eV for
$^{229}$Th:SiO$_2$/Si sample with decaying isomeric $^{229}$Th
nuclei, we substitute it in the  equation $\hbar\omega_N= 5.1 +
KE_{\textrm{max}}$, and obtain the nuclear transition energy of:
\begin{equation*}
E_{\text{is}}=\hbar\omega_N = 7.1(^{+0.1}_{-0.2})~\text{eV}.
\end{equation*}
Comparing to the known value of $7.8 \pm 0.5$~eV obtained in
experiments with a cascade decays \cite{Beck-07,Beck-R}, our
result is 3 times more accurate and deviates by 1.5 joint standard
deviations. Further decrease of uncertainty is possible by direct
optical spectroscopy of VUV photons emitted by isomeric $^{229}$Th
nuclei implanted in the dielectric sample. To  overcome the
problem of weak photon signal one can increase the number of
implanted $^{229}$Th by several orders of magnitude, since the
suggested excitation-implantation method is easily scalable.

\section{The isomeric level half-life and reduced probability of the nuclear transition}
\label{sec:T-B}

The isomeric nuclei decay in $^{229}$Th:SiO$_2$/Si sample shows a
pronounced non-exponential behaviour (Fig.~\ref{fig:Q_e(t)}) which
can be explained by  physical and chemical environment
\cite{Tkalya-18-PRL}. Indeed, the isomeric nuclear decay rate
depends on (i) refractive index surrounding  medium
\cite{Tkalya-00-JETPL,Tkalya-00-PRC} and (ii) presence  of
vacuum/SiO$_2$ and SiO$_2$/Si(001) interfaces which cause
so-called Purcell effect \cite{Purcell-46}. The decay probability
in presence of interfaces is different from the one in infinite
medium by the Purcell factor.

The theory of Purcell effect for the case similar to ours is
discussed in \cite{Chance-75,Chance-78} for the atomic $E1$
transition. We generalized their approach for the case of $M1$
transitions. The formulas for the $M1$ Purcell factors have the
following form \cite{Tkalya-18-PRL}:
\begin{equation}
f_P(z) = f_P^r(z)+f_P^{nr}(z),
\label{eq:f_P}
\end{equation}
where
\begin{equation}
f_P^r(z)= \int_0^{1}{\cal{F}}(z,\kappa)d\kappa,\quad
f_P^{nr}(z)=\int_1^{\infty}{\cal{F}}(z,\kappa)d\kappa,
\label{eq:f_P_r-nr}
\end{equation}
are respectively the radiative and nonradiative decay-rate
constants (see below), $\kappa$ is the reduced transverse momentum
of the photon $\kappa=\sqrt{k^2_{1_x}+k^2_{1_y}}/k_1$, where $k_1
= \sqrt{\varepsilon_1}\omega_N/c$, and $\varepsilon_1$ is the
dielectric constant of the SiO$_2$ film. Function
${\cal{F}}(z,\kappa)$ in Eq.~(\ref{eq:f_P_r-nr}) is defined as
\begin{eqnarray}
{\cal{F}}(z,\kappa) &=& \frac{1}{2}{\text{Im}} \left\{
\frac{F(\hat{d}_0-\hat{z},R_{12}^{\bot})
F(\hat{z},R_{13}^{\bot})}{F(\hat{d}_0,-R_{12}^{\bot}R_{13}^{\bot})}
\frac{\kappa^3}{l_1} \right.  \nonumber\\
&+& \left[(1-\kappa^2) \frac{F(\hat{d}_0-\hat{z},-R_{12}^{\bot})
F(\hat{z},-R_{13}^{\bot})}{F(\hat{d}_0,-R_{12}^{\bot}R_{13}^{\bot})}
\right.  \nonumber\\
&+& \left.\left. \frac{F(\hat{d}_0-\hat{z},-R_{12}^{\|})
F(\hat{z},-R_{13}^{\|})}{F(\hat{d}_0,-R_{12}^{\|}R_{13}^{\|})}
\right]\frac{\kappa}{l_1}\right\}.
\label{eq:F(kappa)}
\end{eqnarray}
Here $F(x,y)=1+y\exp(-2l_1x)$, $\hat{d}_0=k_1d_0$, $\hat{z}=k_1z$
with $d_0$ being the thickness of the SiO$_2$ film ($d_0=6.5$\,nm
in our case). The distance $z$ is counted from the certain
$^{229}$Th nucleus and the vacuum/SiO$_2$ interface (accordingly,
$d_0-z$ is the distance between the $^{229}$Th nucleus and the
SiO$_2$/Si interface). The reflection coefficients in
Eq.~(\ref{eq:f_P}) are defined as
\begin{equation}
R_{1,j}^{\|}=\frac{\varepsilon_1l_j-\varepsilon_jl_1}{\varepsilon_1l_j+\varepsilon_jl_1},
R_{1,j}^{\bot}=\frac{\varepsilon_1-\varepsilon_j}{\varepsilon_1+\varepsilon_j},
\label{eq:R_1j}
\end{equation}
where $j=2,3$, and
$l_j=-i\sqrt{\varepsilon_j/\varepsilon_1-\kappa^2}$ (we remained
notations from \cite{Chance-75,Chance-78}). In
Eqs.~(\ref{eq:f_P})--(\ref{eq:R_1j}), \mbox{$l_1=-i\sqrt{1-\kappa^2}$}.

In Eq.~(\ref{eq:f_P_r-nr}), the function $f_P^{\text{r}}(z)$ gives
the radiative decay-rate constant for energy transfer to the
vacuum through the vacuum/SiO$_2$ interface \cite{Chance-78}.
Correspondingly, the number of emitted photons is proportional to
the function $f_P^{\text{r}}$. The function $f_P^{\text{nr}}(z)$
gives the nonradiative decay-rate constant for energy transfer
into the Si substrate \cite{Chance-78}. This component exists
since the imaginary part of the Si dielectric constant is
different from zero. The number of photoelectrons is
proportional to the energy absorbed in the Si substrate, i.e. to
the function~$f_P^{\text{nr}}$.

\begin{figure}
\includegraphics[angle=0,width=0.90\hsize,keepaspectratio]{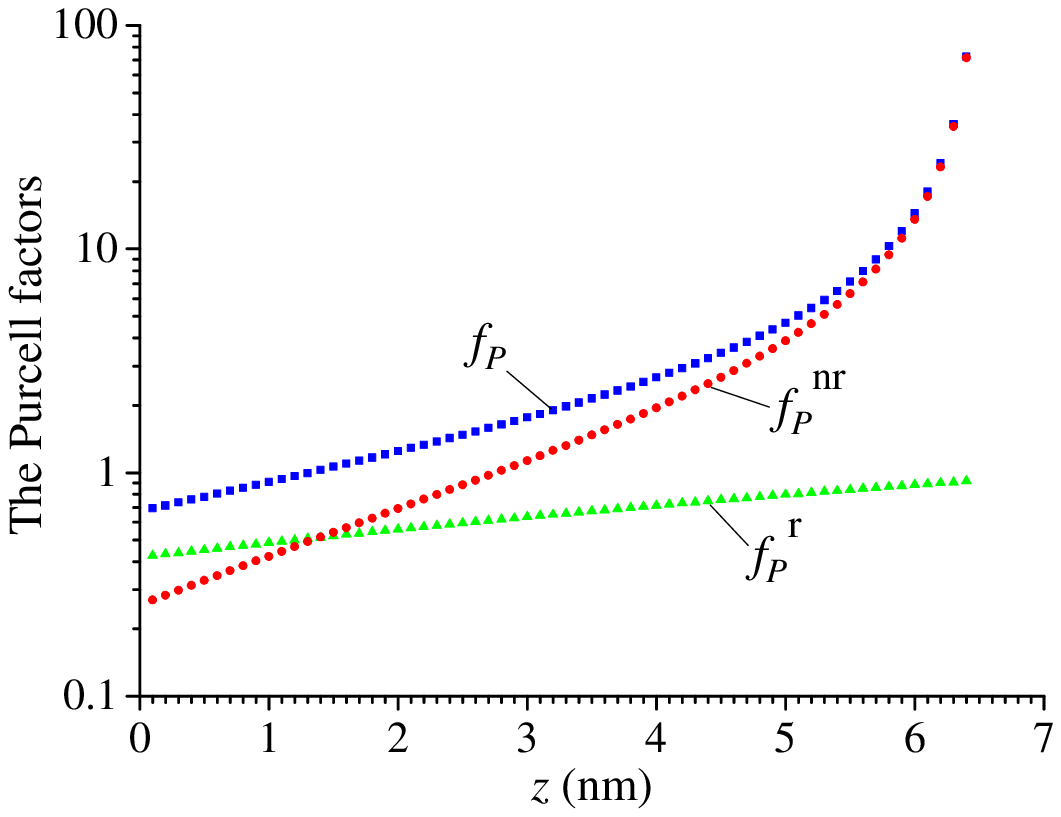}
\caption{The total, $f_P$, radiative, $f_P^{\text{r}}(z)$, and
nonradiative, $f_P^{\text{nr}}(z)$, Purcell factors calculated
from Eqs.~(\ref{eq:f_P}--\ref{eq:F(kappa)}) for the $M1$ isomeric
transition in our $^{229}$Th:SiO$_2$/Si sample for nuclei placed
at the distance $z$ from the vacuum interface.}
\label{fig:PurcellFactor}
\end{figure}

The Purcell factor depends on the position of the emitting object
relative to interfaces. In order to calculate the functions
$f_P(z)$, $f_P^{\text{r}}(z)$, and $f_P^{\text{nr}}(z)$
numerically, we divided thin SiO$_2$ film in $0.1$~nm thick
layers. For each layer at the distance of $z_i$ from the vacuum
interface, the Purcell factor was calculated for the following
values of dielectric constants: $\sqrt{\varepsilon_1} = 1.75$
(measured in this work), $\sqrt{\varepsilon_2} = 0.7+i2.4$ for Si
at the photons energy 7.1 eV \cite{Adachi-99}, and
$\sqrt{\varepsilon_3} = 1$ for vacuum. The result of the
calculation is shown in Fig.~\ref{fig:PurcellFactor}.

%
%

In our experiment, we detect not photons, but photoelectrons from
Si substrate coming from isomeric $^{229}$Th nuclei decay in thin
SiO$_2$ film. The photoelectron signal detected by the
spectrometer (Fig.~\ref{fig:Q_e(t)}) is simulated by the following
function:
\begin{equation}
Q_{\text{e}}(t) = \frac{1}{N} \sum_{i} \lambda_{\text{nr}}(z_i)
e^{-z_i/z_0} e^{- \lambda(z_i) t}, \label{eq:Q_e}
\end{equation}
where $N$ is the normalization providing  $Q_e(0)=1$. Distribution
of $^{229}$Th nuclei in SiO$_2$ along $z$ axis is described by the
exponential function $e^{-z_i/z_0}$ with $z_0 = 6$\,nm
\cite{Borisyuk-18-LPL}. The decay constants $\lambda(z_i)$ and
$\lambda_{\text{nr}}(z_i)$ in $i$-th layer of the SiO$_2$ film are
given by
\begin{eqnarray*}
\lambda(z_i) &=& f_P(z_i)n^3_{\text{SiO}_2}
\ln(2)/T_{1/2},\\
\lambda_{\text{nr}}(z_i) &=&
f_P^{\text{nr}}(z_i)n^3_{\text{SiO}_2} \ln(2)/T_{1/2},
\label{eq:lambda}
\end{eqnarray*}
where  $T_{1/2}$ is the isomeric state half-life of the bare
nuclei in vacuum, $n_{\text{SiO}_2}=\sqrt{\varepsilon_1}$.

Thus, the non-exponential behaviour of the signal shown in
Fig.~\ref{fig:Q_e(t)} can be explained by the fact that isomeric
nuclei implanted close to the SiO$_2$/Si interface have much
faster decay (more than by one order of magnitude) compared to
nuclei sitting close to the Si/vacuum interface. Since the number
density of isomeric nuclei change only by approximately factor of
$e$ from one interface to another, the $z$-dependent Purcell
factor significantly changes dynamics of our signal. It is also
interesting to note that the simultaneous measurement of photons
and electrons would give different decay curves.

We approximate experimental data from Fig.~\ref{fig:Q_e(t)} by
$Q_e(t)$ using only one fit parameter $T_{1/2}$. The result is
\begin{equation}
T_{1/2}=1880\pm170\,\textrm{s} \,
\end{equation}
which corresponds to the half life of the bare isomeric $^{229}$Th
nucleus in vacuum. One can mention very good correspondence
between experimentally measured  and calculated dynamics of the
signal which confirms our model for interaction between excited
nuclei and environment.

From the half-life $T_{1/2}$ one can derive the reduced
probability of the nuclear magnetic-dipole transition from the
excited state with the spin $3/2^+$ to the ground state with the spin
$5/2^+$:
\begin{equation}
B_{W.u.}(M1;3/2^+\rightarrow 5/2^+) = (3.3\pm 0.3)\times 10^{-2}.
\label{eq:Bwu}
\end{equation}
This value is close to the average value of $3.1\times10^{-2}$
obtained in \cite{Tkalya-15-PRC}~(see Table~I) in the frame of the
rotational model.

\section{Appendix I: Experiments with other targets. Cross-checks and discussion }
\label{sec:appI}

We made a set of dedicated experiments aimed to check results of
Section\,\ref{sec:Electron_spectroscopy}. Although there are no
known long-living electronic states in silicon or silicon oxide
which can cause electron emission similar to one shown in
Fig.~\ref{fig:ElectronEnergySpectraThroughSiO2}, it is obligatory
to demonstrate that the signal is caused purely by excited
$^{229}$Th isotope nuclei. We made a number of experiments with
test targets not containing $^{229}$Th isotope. The targets of (i)
metallic $^{232}$Th (isotopically pure), (ii) $^{232}$Th
(isotopically pure) oxide, (iii) graphite, (iv) silicon and (v)
thin silicon oxide film on silicon were laser ablated and then the
regular experimental procedure described in
Section~\ref{sec:Electron_spectroscopy} was carried out. None of
these experiments resulted in any spectral feature similar to the
very peculiar electronic spectra appearing after implantation of
$^{229}$Th isotope
(Fig.~\ref{fig:ElectronEnergySpectraThroughSiO2}).

As a quantitative check we evaluated the expected electron count
rate assuming that it is caused by decay of $^{229}$Th nuclear
isomeric state.

Taking into account expected losses of electrons in our
experimental configuration and knowing the count rate directly
after implantation ($I_{SE}\approx3000$\,s$^{-1}$) the number of
excited nuclei in the sample can be estimated from the relation
\begin{equation*}
N_{\text{is}}=\frac{T_{1/2}}{n^3_{\text{SiO}_2}}\frac{I_{SE}}
{QE_{\text{Si}}\times{} CE\times{} LS}\approx 10^8,
\end{equation*}
where $QE_{\text{Si}}\approx0.3$ is the quantum efficiency of
silicon in VUV range, $CE\approx 0.5$ is the acceptance efficiency
of XSAM-800 spectrometer, $LS\approx 0.17$ is extinction
coefficient of photoelectrons in 6.5\,nm silicon oxide film
(measured with XPS).

Knowing $N_{\text{is}}$, one can evaluate the excitation
probability of $^{229}$Th isomeric nuclear state in ICC process.
Taking into account that the net number of $^{229}$Th nuclei in
the sample is $N_{229}\approx 3\times 10^{12}$, the excitation
probability is on the order $N_{\text{is}}/N_{229}\sim10^{-5}$.
This value is of the same order as theoretical expectation from
equation (\ref{eq:Efficiency}) evaluated for optimal plasma
conditions ($\eta_{\textrm{IIC}}\simeq10^{-5}$,
Eq.~(\ref{eq:Efficiency-Number})).

As follows from Maxwellian distribution, the excitation
probability $\eta_{\textrm{IIC}}$ depends exponentially on the
electron plasma temperature. Thus, the decrease of electron plasma
temperature should result in a sharp decrease of IIC efficiency
and, consequently, the excitation probability. We carried out a
test experiment with decreased  laser power density on the target
 while all other experimental conditions remain
unchanged. In this experiment the illuminated area was increased
by a factor of 4 at the same pulse energy which resulted in 4
times decrease of laser power density. Under this experimental
conditions we did not see any signature of secondary electron
emission which can be ascribed to decay of $^{229}$Th isomeric
nuclear state.

The test experiments clearly indicate that the electron signal is
directly related with implanted $^{229}$Th ions. Moreover, its
amplitude is similar to one expected from ICC process in laser
plasma at optimal parameters and is very sensitive to electron
plasma temperature. We conclude, that the signal comes from decay
of $^{229}$Th low-lying isomeric state as depicted in
Fig.~\ref{fig:229mThDecay-Si-Electron}. Of course, it would be
highly desirable to measure the photon signal directly by optical
methods, but the detection efficiency in current experimental
configuration would be three orders of magnitude lower compared to
electron signal which is close to the noise level coming from
radioactive $^{229}$Th isotope. It should be mentioned, that our
experimental method upgraded with a dedicated optical registration
setup may result in crucial increase of the accuracy for the
transition energy. Even simple narrow-band filters or
low-resolution spectrometer may improve the relative uncertainty
to $10^{-3}$ level, while the high resolution setups allow to
reach $10^{-5}$ relative uncertainty.

It should me noted that the electron spectra from
$^{229}$Th:SiO$_2$/Si sample were measured in the electron
counting regime of XSAM-800 spectrometer. The photoelectron
spectra from VUV light sources were measured in the current
measurement regime (saturation regime). Using one of the VUV
sources we checked that the spectra measured in different regimes
have fully identical shape.

\section{Acknowledgments}

Authors are grateful the following colleagues for helpful
discussions and experimental setup: V.I. Troyan, V.P. Yakovlev,
K.Yu. Khabarova, A.V. Shutov, A.V. Zenkevich, Yu.N. Kolosov.

One of the authors (E.T.) is grateful colleagues from the
University of California, Los Angeles Prof. E.R. Hudson, Dr. Ch.
Schneider, and Dr. J. Jeet for helpful advice and constant
readiness to help in the work.

This research was supported by a grant from the Russian Science
Foundation (Grant No. 16-12-00001).

\end{document}